\documentclass[groupedaddress,superscriptaddress,aps,prd,preprintnumbers,onecolumn,11pt,fleqn,nofootinbib,nobibnote,eqsecnum,floatfix,showpacs,preprint]{revtex4}

\usepackage{graphicx}
\usepackage{bm,amsmath,amssymb}
\usepackage[mathscr]{eucal} 

\long\def\comment#1{ }
\newcommand{\eqn}[1]{Eq.~\eqref{#1}}
\newcommand{\beq}{\begin{equation}}
\newcommand{\eeq}{\end{equation}}
\newcommand{\bea}{\begin{eqnarray}}
\newcommand{\eea}{\end{eqnarray}}
\newcommand{\nn}{\nonumber\\}
\newcommand{\dif}{{\rm d}}

\newcommand{\rmi}{{\rm i}}
\newcommand{\rmP}{{\rm P}}

\newcommand{\rmTr}{{\rm Tr}}
\newcommand{\del}{\partial}

\newcommand{\mcal}{\mathcal}

\newcommand{\wh}{\widehat}

\newcommand{\bp}{\bm{p}}
\newcommand{\bx}{\bm{x}}

\newcommand{\abar}{\bar{\alpha}_s}

\def\s#1{{\scriptscriptstyle #1}}

\newcommand{\deltam}{\delta\hspace{-.025cm}}

\def\fig#1{Fig.~\ref{#1}}

\def\1eq#1{Eq.~(\ref{#1})}

\def\2eqs#1#2{Eqs.~(\ref{#1}) and~(\ref{#2})}
\def\3eqs#1#2#3{Eqs.~(\ref{#1}),~(\ref{#2}) and~(\ref{#3})}
\def\noeq#1{(\ref{#1})}

\def\s#1{{\scriptscriptstyle #1}}

\def\c{C}
\def\bc{\bar{C}}
\def\b{B}
\def\csf{c}
\def\bcsf{\bar\csf}
\def\bsf{b}
\def\cso{\delta c}
\def\bcso{\delta \bar c}
\def\bso{\delta b}

\def\t{\theta}

\def\G{\Gamma}

\begin{document}

\title{High-energy QCD evolution from BRST symmetry}

\author{D.~Binosi}
\email{binosi@ectstar.eu}
\affiliation{European Centre for Theoretical Studies in Nuclear Physics and Related Areas (ECT*)\\ 
and Fondazione Bruno Kessler, Strada delle Tabarelle 286, I-38123 Villazzano (TN), Italy}

\author{A.~Quadri}
\email{andrea.quadri@mi.infn.it}
\affiliation{Dip.~di Fisica, Universit\`a degli Studi di Milano, Via Celoria 16, I-20133 Milano, Italy\\
and INFN, Sezione di Milano, Via Celoria 16, I-20133 Milano, Italy}

\author{D.N.~Triantafyllopoulos}
\email{trianta@ectstar.eu}
\affiliation{European Centre for Theoretical Studies in Nuclear Physics and Related Areas (ECT*)\\
and Fondazione Bruno Kessler, Strada delle Tabarelle 286, I-38123 Villazzano (TN), Italy}

\date{February 11, 2014}

\begin{abstract}

We show that the (gauge fixed) classical action of the Color Glass Condensate is invariant under a suitable Becchi-Rouet-Stora-Tyutin symmetry, that holds after the gluon modes are split into their fast, semi-fast and soft components, according to the longitudinal momenta they carry. This entails the existence of a corresponding Slavnov-Taylor identity which in turn strongly constrains the effective field theory arising when integrating out the semi-fast modes. Specifically, we prove that this identity guarantees the gauge invariance of the resulting effective theory. In addition, we use it to demonstrate that the integration over the semi-fast modes does not deform the classical Yang-Mills equations of motion, thus validating a key assumption in the usual procedure adopted when  deriving the renormalization group equation governing the evolution with energy of the effective theory. As far as the latter are concerned, we finally prove that its functional form is common, and it is determined by symmetries arguments alone. The formal properties of these equations valid in different regimes and/or approximations 
({\it e.g.},~the JIMWLK equation and its BFKL limit) can be therefore derived in a unified setting within this algebraic approach. 

\end{abstract}


\maketitle


\section{\label{sec:intro}Introduction and motivation}

The physics of high gluon densities and gluon saturation is one of the subfields of Quantum Chromodynamics (QCD) which has continuously attracted much attention over the last years, both theoretically and experimentally. The initial interest was concentrated on electron-proton deep inelastic scattering, but recently it has been shifted to the study of heavy ion collisions \cite{Iancu:2012xa,Kovchegov:2012mbw}, as gluon saturation plays a critical role in describing the initial wavefunction of the colliding nuclei and the early stages of the collision towards thermalization.

The first QCD based calculation at small-$x$, where by $x$ we refer to the longitudinal momentum fraction of a parton, resulted in the BFKL equation \cite{Lipatov:1976zz,Kuraev:1977fs,Balitsky:1978ic}, which predicted a sharp rise of the gluon distribution in the limit of interest. After the seminal work \cite{Gribov:1984tu} in which the idea of gluon saturation was introduced and its necessity was emphasized, and a complementary attempt \cite{Bartels:1980pe} based on imposing unitarity constraints, various methods to address physics at small-$x$ were developed. Here we shall only deal with the Color Glass Condensate (CGC), a modern approach which is based on effective field theory (EFT) techniques for integrating out degrees of freedom, and provides a well-defined framework that can be used for phenomenological applications \cite{Albacete:2013tpa}.

The main concepts of the CGC were already contained in the so-called McLerran Venugopalan (MV) model~\cite{McLerran:1993ni,McLerran:1993ka},  which aimed at describing the gluon distribution at small-$x$ in a very large nucleus, that is, with atomic number $A \gg 1$. For transverse separations smaller than $1/\Lambda_{\rm QCD}$ this model treats the $A \times N_c$ valence quarks as uncorrelated long-lived color sources for the emission of soft gluons. Due to the large number of nucleons, a strong coherent color field can be created that leads to the saturation of gluon occupation numbers which become of order $1/\alpha_s$. Despite its simplicity, since it does not contain any quantum (chromo)dynamics, the MV-model, or at least some refined versions of it, is still a reliable model for providing the initial conditions at moderate values of $x$ in heavy ion collisions.

However, such dynamics is necessary in order to evolve the wavefunction of a hadron (or a nucleus) to arbitrarily small values of $x$. This is not such an easy task as it requires the resummation of large longitudinal logarithms in a dense environment. This program was quite successful and resulted in a renormalization group (RG) equation, known as the JIMWLK equation\footnote{The acronym stands for Jalilian-Marian, Iancu, McLerran, Weigert, Leonidov and Kovner.} \cite{JalilianMarian:1997jx,JalilianMarian:1997gr,Kovner:2000pt,Iancu:2000hn,Ferreiro:2001qy,Iancu:2001ad,Weigert:2000gi}. This is a functional Fokker-Planck equation for the evolution of a weight-function $W[\rho]$ which determines all the correlations of the color sources $\rho$. For a certain observable, the scattering of a small color dipole off the CGC in the multi-color limit, the JIMWLK equation leads to the Balitsky-Kovchegov (BK) equation \cite{Balitsky:1995ub,Kovchegov:1999yj}, which is a closed non-linear evolution equation. To a large extent, the solution to the JIMWLK equation has by now been understood \cite{Dumitru:2011vk,Iancu:2011ns,Iancu:2011nj} and, being a Fokker-Planck equation, it has an equivalent Langevin formulation \cite{Blaizot:2002xy} which has been recently extended to address the problem of gluon correlations at different values of $x$ \cite{Iancu:2013uva}. Let us also note that various works have appeared after the final version of the equation was written, in which simpler derivations have been given, certain aspects have been clarified or the validity of the equation has been extended to larger kinematic regimes \cite{Mueller:2001uk,Kovner:2005nq,Hatta:2005rn,Hatta:2005wp,Fukushima:2006cj,Gelis:2008rw,Jeon:2013zga,Caron-Huot:2013fea,Balitsky:2013fea,Kovner:2013ona}.

As said, the derivation of the JIMWLK equation is rather involved, since one has to resum longitudinally enhanced contributions in the presence of a potentially strong background field. Moreover it requires a special blend of gauge choices, mostly based on physical intuition, where the background field is kept in the Coulomb gauge while the modes to be integrated over are in the light-cone gauge.  Here we would like to use techniques which are exploiting the Becchi-Rouet-Stora-Tyutin (BRST) symmetry and the  associated (extended) Slavnov-Taylor (ST) identity in the presence of a non-trivial background \cite{Binosi:2011ar,Binosi:2012pd,Binosi:2012st}, in order to disentangle the gauge-dependent effects, due to the
specific gauge choice adopted, from the gauge-invariant physical quantities of the EFT.
In particular, we will show that gauge-invariance of the CGC effective action holds as a consequence of the fulfillment of the ST identity (after the integration of the semi-fast modes); hence, as expected on general physical grounds, the specific gauge used during the one step quantum evolution is only a matter of convenience\footnote{For example, if one uses a renormalizable gauge as opposed to the Light Cone gauge that is usually conveniently employed for the semi-fast modes in most of the CGC calculations, the derivation becomes more complicated (ghosts cannot be neglected anymore) but the final results are granted to coincide.}.
We hasten to emphasize that this proof is not limited to the one-loop approximation, but holds to all orders in the perturbative expansion
on the semi-fast modes; in addition, being based on symmetry arguments only, it is regularization scheme-independent as well (as far as the ST identity is not broken).

Moreover, one can also study how the background field equation of
motion changes once quantum corrections are taken into account.
Such an equation of motion is completely fixed by
the  ST identity in the presence of a non-trivial background
and can be solved by a certain canonical transformation~\cite{Binosi:2012st,Binosi:2012pd}. In the CGC approach the equation of motion for the
background fast modes, valid after the one-step quantum
evolution, is crucial in deriving the CGC evolution equations, since
one must be able to prove that the updated background configuration
can again be obtained by the same classical Yang-Mills
equation, now in the presence of color charges with new correlators
encoding the effects of the integration of the semi-fast gluons.
That this is indeed the case is far from obvious,
since, in general, the classical background equation of motion
is not preserved once quantum corrections are taken into account~\cite{Binosi:2011ar}. Still, as we will show, in the CGC approach the clever choice of the (background)
gauge and of the definition of the color charges~\cite{Hatta:2005rn}
stabilizes the classical background equation of motion
under radiative corrections, thus ensuring the formal consistency
of the whole picture. 

The structure of the paper is as follows. In Sect.~\ref{sec:cgc} we review the CGC, paying particular attention to the physical motivation for constructing such an EFT and introducing the appropriate action and its symmetries. In Sect.~\ref{sec:brst} we discuss the BRST transformations of the various fields appearing in the action and in Sect.~\ref{sec:gf} we elaborate on the gauge fixing term.
In Sect.~\ref{sec:sti}, which may be considered as the main section of the paper, 
we obtain the ST identity for the CGC effective action,
generated upon the integration over 
the semi-fast modes introduced in Sect.~\ref{sec:cgc}.
The ST identity imposes strong constraints on the CGC effective action
for the soft gluon modes (in the presence of the fast background). Specifically, we will show that it is the
key tool for establishing two important properties: i) the gauge invariance
of the effective action (irrespectively of the choice of the
gauge-fixing adopted in the integration of the semi-fast modes which is kept unspecified),
to be discussed in Sect.~\ref{sec:gaugeinv}, and ii) the stability of the
quantum-corrected equation of motion for the background derived in Sect.~\ref{sec:qcor}.
In Sect.~\ref{sec:def} we will indeed show that the classical field is not deformed by the one-step quantum evolution and that the classical relation between the 
background field and the color sources remains true after integrating the semi-fast modes. In Sect.~\ref{sec:evol} we show how the color charge correlations, generated from the quantum evolution, can be obtained from the effective action,
leading to the CGC evolution equations.
The general pattern of the derivation is dictated only
by symmetry arguments, while the explicit expressions of the 
evolution Hamiltonian of course depends on the 
particular approximation used in the computation of the effective action.
Conclusions are presented in Sect.~\ref{sec:concl}, while
Appendix~\ref{app:jimwlk} contains a sketch of the derivation of the JIMWLK equation.

\section{\label{sec:cgc}The Color Glass Condensate}

A generic hadron or a nucleus in its rest frame is a rather complicated object. It contains nucleons (in the case of a nucleus), a few valence quarks and zillions of quarks and gluons, all of which are confined to live in the space occupied by the hadron\footnote{Here we use the term hadron to also include the case of a nucleus with an arbitrary atomic number.} under question. The typical time scale for the strong interactions among the hadron constituents is $\rm{1/\Lambda_\s{\rm QCD}}$, since there is no other scale in the problem, and therefore one cannot say much without relying on non-perturbative methods.

The above description changes drastically when we go to the infinite momentum frame, a frame in which the hadron moves ultra-relativistically, usually along the $x^3$ direction by convention. Then, hadronic time scales are dilated by a large Lorentz factor $\gamma$ and one has the possibility to separate calculable, but non-trivial, perturbative QCD dynamics from non-perturbative ones, as for example done in the standard analysis of electron-proton deep inelastic scattering.

At high-energy, by definition, we are interested in kinematics such that the hadron's longitudinal momentum is much larger than all possible transverse momenta, with the latter assumed to be larger than $\Lambda_\s{\rm QCD}$ so that we can rely on weak coupling techniques. For example, and in order to be more pragmatic, it has to be much larger than the transverse momenta of produced particles when collided with another hadron. In such a kinematic regime, a prominent role is played by the small-$x$ gluons, which are those gluons carrying a small fraction $x$ of the hadron's total longitudinal momentum.

QCD favors the generation of such small-$x$ gluons, since the emission of a gluon (but not that of a quark) with 
fraction $x$ from a parton (either quark or gluon) with fraction $x_0$ is proportional to $\abar \ln(x_0/x)$, with $\abar = \alpha_s N_c/\pi$, $\alpha_s = g^2/4 \pi$ the QCD coupling and $N_c$ the number of colors. Clearly, in the limit of interest, the logarithm can overcome the smallness of $\abar$ and one needs to resum powers of $\abar \ln(x_0/x)$ to all orders in perturbation theory. This is equivalent to viewing this slowest gluon with fraction $x$ as being the lower end of a cascade composed of $n$ successive intermediate emissions of gluons with strongly ordered longitudinal momentum fractions, that is with $x_0 \gg x_1 \gg \dots \gg x_n \gg x$. On the contrary, transverse momenta are not ordered and therefore the transverse sizes of gluons remain typically the same in the course of evolution towards smaller values of $x$. Then, the aforementioned resummation of the perturbative series leads to a fast, exponential in the rapidity $Y\equiv \ln(1/x)$, increase in the gluon occupation number, {\it i.e.},~in the number of gluons per unit phase space. This violates unitarity, since an occupation number should not be larger than $\sim 1/\alpha_s$.

What has gone wrong in the above picture, is that we have assumed small-$x$ gluons to be emitted independently from its predecessors, an assumption which is well-justified so long as the wavefunction is still dilute and naturally leads to exponentiation. However, this is not true any more when occupation numbers grow large. Then the emission of a small-$x$ gluon is a coherent phenomenon as the gluon is subject to live in a dense environment. This mechanism suppresses the emission rate which eventually saturates, leading to (marginal) saturation\footnote{More precisely, the gluon occupation number stil grows, albeit very slowly, like $\sim \ln (1/x)$.} of the gluon occupation number consistent with field theoretical requirements. 

The Color Glass Condensate is a modern effective field theory which encompasses the above description for the small-$x$ components of the ultra-relativistic hadronic wavefunction. ``Color'' stands for the fact that we deal with a Yang-Mills theory and ``condensate'' is for the occupation numbers which can reach values of order $1/\alpha_s$. The characterization ``glass'' comes about because gluons with different longitudinal momenta have different lifetimes. To be more precise, let us first introduce the light-cone coordinates $x^{\mu} = (x^{+},x^{-},\bx)$ with $x^{\pm} = (x^0\pm x^3)/\sqrt{2}$ and $\bx=(x^1,x^2)$. For our convenience later on, let us also define here the 3-vector $\vec{x} = (x^-,\bx)$. Then the lifetime of a gluon is $\tau \sim p^+/\bp^2 = xP^{+}/\bp^2$, where $P^+$ is the longitudinal momentum of the hadron, meaning that gluons with smaller-$x$ live shorter.  Thus, a gluon with fraction $x$ sees all the partons from which it has been coherently emitted as static, {\it i.e.},~as $x^+$-independent, color sources. Moreover, by the same token, since these sources have much larger longitudinal momenta, they also have much shorter longitudinal wavelengths and therefore they appear to the emitted gluon as sharply localized in $x^-$.

So, let us consider an arbitrary longitudinal scale $\Lambda$ (clearly not to be confused with $\Lambda_\s{\rm QCD}$). If we are interested in correlations of ``slow'' gluons with momenta $k^+$ smaller, but not much smaller, than $\Lambda$, one can integrate all the QCD dynamics above $\Lambda$ and absorb them in the aforementioned ``fast'' static color sources with a charge density $\rho^a(x^-, \bx)$. Of course these color sources have, in principle, highly non-trivial correlations which can be conveniently summarized in the weight-function $W_{\Lambda}[\rho]$, which is a functional probability distribution. These correlations are automatically transmitted in the correlations of gluons with momenta below, but not very far from, the scale $\Lambda$. Thus, recalling also that we are interested in potentially large occupation numbers, or equivalently large gauge fields of order $1/g$, we see that our problem turns into a classical Yang-Mills theory in the presence of sources. 

As said, the weight-function $W_{\Lambda}[\rho]$ includes all the   quantum dynamics, among which non-perturbative effects, so that eventually one will have to resort to a modelling of infrared physics. However, one can predict how $W_{\Lambda}[\rho]$ evolves, and the resulting evolution is perturbative as we will explain below. If we become interested in even lower momenta $p^+ \sim b\Lambda$ with $b \ll 1$, then it becomes obvious that some of the modes which were soft, now  become fast and  have to be integrated in the sources. Thus, the correlations of these color sources get logarithmically enhanced contributions of order $\abar \ln(\Lambda/p^+) \simeq \abar \ln(1/b)$, and if the whole construction scheme is correct, these contributions should be absorbed in a new weight-function $W_{b\Lambda}[\rho]$. 

Clearly, one may wonder why this calculation is infrared safe. The straightforward answer is that this is done at the leading logarithmic level and the QCD coupling is taken to be fixed. Still, this is not a satisfactory answer because if the evolution becomes sensitive to very small transverse momenta, then it is almost guaranteed that the next to leading calculation will suffer from infrared divergences. However, the saturation of occupation numbers simply says that there is a scale $Q_s$, called the saturation momentum (or saturation scale), below which the initial exponential growth in $Y=\ln(1/x)$ is tamed. This scale, which is dynamically generated, is a perturbative one as it increases rapidly with $Y$ \cite{Gribov:1984tu,Iancu:2002tr,Mueller:2002zm,Triantafyllopoulos:2002nz,Munier:2003vc}, which means that even modes with arbitrarily high $\bp$ will saturate at sufficiently small-$x$. Therefore, $Q_s$ is the natural scale to set the value of the coupling and moreover saturation (in the form of non-linear terms in evolution equations) will cut potentially dangerous infrared contributions in the course of evolution.

After this introductory description we come to the level where we can formulate our problem. Let us start by writing the action of our theory which reads
 \beq
 \label{action}
 S_\s{\rm CGC}[A,\rho] = S_\s{\rm YM}[A] + S_\s{W}[A,\rho].
 \eeq
In the expression above $S_\s{\rm YM}$ is the Yang-Mills action
 \beq
 S_\s{\rm YM} = -\int \dif^4 x\,\frac{1}{N_c}\, \rmTr[F_{\mu\nu} F^{\mu\nu}],
 \eeq
where the field strength is given in matrix form $F_{\mu\nu} = F_{\mu\nu}^a T^a$, with $T^{a}$ the generators in the adjoint representation of the SU($N$) group; in components we have $F_{\mu\nu}^a = \del_{\mu} A_{\nu}^a - \del_{\nu} A_{\mu}^a + g f^{abc} A_{\mu}^b A_{\nu}^c$ with $f^{abc}$ the SU($N$) structure constants. 

The piece $S_\s W$ in \eqn{action} contains the interactions of the color source $\rho$ which stands for the plus component (the only non-vanishing one) of the 4-current associated with the fast sources. These sources couple to the $A^-$ component of the gauge field and as a first attempt, one may guess that $S_\s W$ is proportional to $\rho^a(\vec{x}) A_a^-(x)$. Such a term though cannot be gauge invariant, and eventually one has to define the action on a Schwinger-Keldysh contour in the complex time plane. It is given by \cite{Iancu:2000hn}
 \beq
 \label{swc}
 S_\s{W}[A,\rho] = \frac{\rmi}{g N_c}
 \int \dif^3 \vec{x}\, \rmTr\big[\rho(\vec{x})\, W_\s{C}(\vec{x})\big],
 \eeq
where $W_\s C(\vec{x})$ is the contour temporal Wilson line
 \beq
 \label{wc}
 W_\s{C}(\vec{x}) = {\rm T}_\s{C}\exp \left[  \rmi g \int_C \dif z\, A^-(z,\vec{x}) \right].
 \eeq
The contour $C = C_+ \cup C_-$ in Eqs.~\eqref{swc} and \eqref{wc} is the aforementioned Schwinger-Keldysh contour defined as follows: $C_+$ is the path along the real time axis, from $x_0^+$ to $x_f^+$, while the points on $C_-$ have a small imaginary part, that is $z = x^+ - \rmi \eta$ with $\eta \to 0_+$, and $x^+$ runs backwards from $x_f^+$ to $x_0^+$. Eventually we shall take the limits $x_0^+ \to -\infty$ and $x_f^+ \to +\infty$. In \eqn{wc} ${\rm T}_\s C$ orders the matrices $A^-$ from right to left as ones moves along the contour $C$, {\it i.e.},~it coincides with chronological ordering along $C_+$ and anti-chronological ordering along $C_-$.

Let us now assume that $G(x) \in {\rm SU}(N)$ satisfies for any $\vec{x}$ the periodic condition
 \beq
 \label{omega}
 G(-\infty - \rmi \eta , \vec{x}) = G(-\infty , \vec{x}).
 \eeq
Due to this property, one easily sees that the gauge transformations
 \begin{align}
 \label{gaugea}
 & A^{\mu}(x) \,\mapsto\, 
 G\, A^{\mu}(x)\, 
 G^{\dagger}(x)
 + (\rmi/g)\,
 G(x)\, \del^{\mu}\, G^{\dagger}(x)
 \\
 \label{gaugerho}
 & \rho(\vec{x}) \mapsto
 G(-\infty,\vec{x})\,
 \rho(\vec{x})\,
 G^{\dagger}(-\infty,\vec{x})
 \\
 \label{gaugew}
 & W_\s{C}(\vec{x}) \mapsto
 G(-\infty - \rmi \eta, \vec{x})\,
 W_\s{C}(\vec{x})\,
 G^{\dagger}(-\infty,\vec{x}),
 \end{align}
leaves the action $S_\s{W}[A,\rho]$ invariant. It is also instructive to notice that $S_\s{W}[A,\rho]$ may be equivalently written as
 \beq
 \label{sw2}
 S_\s{W}[A,\rho] = - \frac{1}{N_c}
 \int_C \dif^4 x \, \rmTr \big[ \rho(\vec{x})\, A^{-}(z,\vec{x})\, W_{z,-\infty}(\vec{x})\big],
 \eeq 
where the subscripts in the Wilson line simply mean that the contour integral in the complex time plane should now start at $-\infty$ (on the upper branch) and end at $z$ (either in the upper or in the lower branch). The lowest order term of the Wilson line in \eqn{sw2} leads to the linear (in the field) coupling proportional to $\rho^a(\vec{x}) A_a^-(x)$, while higher order terms restore the gauge invariance of the interaction.

Thus, having built a gauge invariant action, one is guaranteed to get the proper classical equations of motion in the presence of a color source. Differentiating the action with respect to (w.r.t.)~$A_a^{-}$ (since this is the only component which couples to the source and thus modifies the classical Yang-Mills equations in the vacuum) we arrive at
 \beq
 \label{maxrho}
 \mcal{D}_{\nu}[A] F^{\nu\mu}(x) = 
 \delta^{\mu+}
 W_{x^+,-\infty}(x)\, \rho(\vec{x}) \, W^{\dagger}_{x^+,-\infty}(x)
 \equiv J^{\mu}(x), 
 \eeq
where we have introduced the covariant derivative
 \beq 
 \label{dmat}
 \mcal{D}_{\nu}[A] \Phi = \del_{\nu} \Phi - \rmi g [A,\Phi],
 \eeq
for a generic matrix field $\Phi$. Thus, the r.h.s.~of \eqn{maxrho} means that the source $\rho$ is subjected to a color precession due to eikonal scattering off the time-dependent field $A^-$. This color precession is necessary in order to have covariant conservation of the current $J^{\mu}(x)$, that is $\mcal{D}_{\mu}[A]J^{\mu}(x) = 0$.
 
In the absence of $A^{-}$, the source becomes $x^+$-independent, $F^{-i}$ (with $i=1,2$) vanishes automatically and it is possible to construct a solution with $F^{ij}=0$, that is $A^i$ is a pure gauge. Then, the only non-trivial field strength component is $F^{+-}$. Choosing a gauge finally fixes $A_i$, leaving us with only one independent degree of freedom, $A^+$; in particular, in the Coulomb gauge ($\del_i A^i = 0$), the classical equation of motion reduces to the Poisson equation
 \beq
 \label{poisson}
 -\bm{\nabla}_{\bx}^2A^{+}(\vec{x})= \rho(\vec{x}).
 \eeq
Then, by a gauge rotation, it is possible to obtain the classical solution in an arbitrary gauge.

Returning now to the quantum problem, one observes that the fact that we are in the infinite momentum frame makes possible to identify the quantum modes to be integrated out when performing one step in the quantum evolution, by splitting the gauge field $A^\mu$ into three pieces according to their support in momentum space. Specifically, we set
 \beq
 A_{\mu} = \wh{A}_{\mu} + a_{\mu} + \deltam A_{\mu},
 \label{asplit} 
\eeq
where (i) $\wh{A}_{\mu}$ represents the fast modes with longitudinal momenta $|p^+| > \Lambda$ and is given by the classical solution to \eqn{maxrho} in the absence of $A^-$ and in an arbitrary gauge as it has been described  above, (ii) $a_{\mu}$ stands for the semi-fast modes, {\it i.e.},~the modes with momenta $p^+$ such that $\Lambda > |p^+| > b \Lambda$ (where we recall that $b \ll 1$ but with $\abar \ln(1/b) \ll 1$) which will be integrated in the one-step quantum evolution, and (iii) $\delta A_{\mu}$ corresponds to the soft modes with momenta $|p^+| < b \Lambda$ which, like $\wh{A}_{\mu}$, will be kept fixed during the quantum step. 

\section{\label{sec:brst}BRST transformations}

We now come to study the BRST symmetry of the action and the corresponding transformations of fields and sources. We will first start from the gluon and ghost sector, while later on we will focus on the classical color charge.

\subsection{Gluons}

When considering the total gauge field $A_{\mu}^a$, its corresponding BRST transformation coincides with the conventional gauge transformation in which the gauge parameter is replaced by the 
associated Faddeev-Popov ghost field~$\c^a$, that is one has (with $s$ the BRST operator)
 \beq
 \label{sona}
 s A_{\mu}^a = \mcal{D}_{\mu}^{ab}[A]\c^b;
 \qquad
 \mcal{D}_{\mu}^{ab}[A] = \delta^{ab} \del_{\mu} - g f^{abc} A_{\mu}^{c},
 \eeq
where $s$ represents the BRST operator, and we have written the covariant derivative defined in \eqn{dmat} in component form.
Next, for the background field $\wh{A}^a_{\mu}$, one introduces the source $\Omega^a_{\mu}$ as its BRST doublet partner through\footnote{Briefly, a pair of variables $(u,v)$ such that $su=v$, $s v =0$ is called a BRST doublet (with $v$ representing the BRST doublet partner of $u$). In ordinary perturbative quantum field theory, Eq.~(\ref{sonahat}) implements the so-called doublet mechanism~\cite{Piguet:1995er,Barnich:2000zw,Quadri:2002nh},  preventing the background field from modifying the physical observables of the model.}~\cite{Grassi:1995wr,Ferrari:2000yp,Becchi:1999ir}  
 \beq
 \label{sonahat}
 s \wh{A}^a_{\mu} = \Omega^a_{\mu};
 \qquad 
 s \Omega^a_{\mu} = s^2 \wh{A}^a_{\mu} =0.
 \eeq  

In the standard formulation of the background field method, where one has the decomposition $A^a_\mu=\wh{A}^a_\mu+Q^a_\mu$,  the transformations~\eqref{sonahat} and \eqref{sona} would have allowed the unequivocal determination of the BRST transformation of the quantum field~$Q^a_\mu$. However, in the present case there is an ambiguity due to the fact that the field $Q^a_\mu$ comprises two terms:  the semi-fast modes $a^a_{\mu}$ (to be integrated over), and the soft modes $\deltam A_{\mu}^a$.  

It turns out that there is no way to disentangle the individual transformations of these two contributions other than resorting to a physical argument of some kind. In this case the latter is provided by the fact that in order to preserve the BRST invariance of the action $S_\s{W}[\delta A,\rho]$,
$\delta A_{\mu}^a$ must clearly transform as a gauge connection.
In addition, since in the EFT spirit one is interested in the Green's functions obtained after the semi-fast modes $a^a_\mu$ are integrated out, it is also natural to split the ghost field $\c^a$ into a soft and a semi-fast component:
\beq
\c^a =\csf^a  +  \cso^a
\label{c.split}
\eeq
where as in the gauge field case, $\cso^a$ denotes the soft modes and $\csf^a$ the semi-fast modes.
Then one demands that
\beq
s\, \deltam A^a_\mu = \partial_\mu \cso^a + g f^{abc} \deltam A^b_\mu \cso^c ,
\label{BRST.deltaA}
\eeq
thus implementing the requirement that the soft field $\delta A^a_\mu$ transforms as a gauge connection w.r.t. the soft ghost $\cso^a$.

At this  point, the transformation of $a^a_\mu$ is fixed by 
the BRST variation of $A^a_\mu$ in~\1eq{sona} and by \1eq{sonahat},
once the splitting of the gluon and ghost fields of~\2eqs{asplit}{c.split} is imposed:
\begin{align}
s a^a_\mu & = s A^a_\mu - s \delta A^a_\mu - s \wh{A}^a_\mu \nonumber \\
         & = \partial_\mu \csf^a + g f^{abc} \left( \wh{A}^b_\mu+ a^b_\mu + \delta A^b_\mu \right) \csf^c  + g f^{abc}\left( \wh{A}^b_\mu+ a^b_\mu\right) \cso^c - \Omega^a_\mu .
\label{BRST.semifast}
\end{align} 

\subsection{Color charge}

In the presence of $\delta A^-$  the current $J^\mu$ appearing on the r.h.s. of~\1eq{maxrho} is evaluated by using the  temporal Wilson line from
$z^+ \rightarrow -\infty$ to $x^+$ of the soft modes $\delta A^-$:
\beq
W_{x^+,-\infty}(x) = \rmP \exp \left[ ig \int_{-\infty}^{x^+} \mathrm{d}z^+\,
\delta A^-(z^+,{\bf x}) \right].
\label{Wwilsonline}
\eeq
The current $J^\mu$ is then covariantly conserved and expresses the color precession of the static color charges in the presence of the time-dependent
fields $\delta A^-$.

The classical solution $\wh{A}^a_\mu$ is determined
by the time-dependent generalization of the solution in \eqn{poisson}, that is from 
 \beq
 \label{rhohat}
 \bm{\nabla}_{\bx}^2 \alpha(x) = - U^\dagger(x) J^+(x) U(x),
 \eeq 
with 
$J^+$ the plus component of the color-rotated current
in the r.h.s. of \1eq{maxrho}, and
$U$, $U^\dagger$ the Wilson lines defined according to
 \beq
 \label{wilsonu}
 U^{\dagger}(x) = 
 \rmP \exp \left[ \rmi g \int_{-\infty}^{x^-}
 \dif z^- \alpha(x^+,z^-,\bx) 
 \right], 
 \eeq
where $\alpha(x)$ is to be identified with the only non-zero component of the classical background field in the Coulomb gauge, that is $\alpha(x)=\wh{A}^+(x)$.

It is most convenient to work directly with the color charge entering in the r.h.s. of \1eq{rhohat} by setting
\beq
\label{chi-charge}
\chi(x) \equiv  U^\dagger(x) J^+(x) U(x).
\eeq
Then, $\chi$ becomes the independent variable and the original charge density $\rho=\rho(\chi)$ is determined by inverting~\1eq{chi-charge} above; inserting the resulting expression in~\1eq{sw2} yields finally $S_\s{W}[A,\chi]$. The BRST transformation of $\chi$ is easily derived after observing that
by~\1eq{chi-charge} $\chi$ transforms in the adjoint representation of SU(N), and therefore
\beq
\label{BRST.chi}
s \chi^a = g f^{abc} \chi^b \cso^c.
\eeq
Notice that the Wilson line $W$ in Eq.~(\ref{Wwilsonline})
only depends on $\delta A$ (and not on $a$), so that the 
BRST transformation in Eq.(\ref{BRST.chi}) contains only 
the soft ghosts $\cso$.

To take into account the fluctuations in the color charge density $\chi^a$ induced by the integration over the semi-fast gluons $a$, we next split $\chi^a$ according to
 \beq
 \label{chisplit}
 \chi^a = \wh{\chi}^{\,a} + \deltam \chi^a.
 \eeq 
Here, $\wh{\chi}^{\,a}$ coincides with the current generating the classical
configuration $\alpha$ in~\1eq{rhohat}, and the corresponding  BRST transformation  can be therefore read off directly from its defining equation supplemented with~\1eq{sonahat}:
\beq
\label{sonwhchi}
s \wh{\chi}^a = - \bm{\nabla}_{\bx}^2 \Omega^{a+}(x).
\eeq
The charge density $\delta\chi^a$ corresponds instead to that  of the semi-fast gluons, and its  BRST transformation can be finally obtained from the splitting~(\ref{chisplit}) 
 \beq
 s\, \deltam \chi^a(x) = s \chi^a(x) - s \wh{\chi}^{\,a}(x),
\label{sondchi}
 \eeq
with the r.h.s. determined by~\2eqs{BRST.chi}{sonwhchi}. The important aspect to notice is that this transformation is independent of the soft gluon field~$\deltam A$.

\section{Gauge fixing\label{sec:gf}}

The derivation of evolution equations such as
the JIMWLK equation or its BFKL limit, requires to integrate out the semi-fast quantum fluctuations. It is at this point that the flexibility of working in the background field formulation of the theory manifests itself, as one has the freedom of choosing different gauge fixings for background and quantum fields (that is, the semi-fast and soft modes in the case at hand).
In  momentum space representation, this can be achieved by choosing a gauge fixing functional of the type
\beq
{\cal F}^a(p) = \t(|p^+| - b \Lambda) \t(\Lambda - |p^+|){\cal F}^a_{\mathrm{s.fast}}(p) + \t(b \Lambda - |p^+|){\cal F}^a_{\mathrm{soft}}(p),  
\label{gf.funct}
\eeq
where the semi-fast (soft) gauge fixing function ${\cal F}_{\mathrm{s.fast}}$ (${\cal F}_{\mathrm{soft}}$) depends only on the semi-fast (soft) gluon modes $a$ ($\delta A$).  The resulting gauge-fixing action can be then calculated through the
usual formula 
\beq
S_\s{\rm GF + FPG} = \int\! \dif^4x \, 
s \left( \bar \c^a {\cal F}^a + \frac{\xi}{2} \bar \c^a B^a \right),
\label{sgf}
\eeq
where $\xi$ is a gauge-fixing parameter, $\bc^a$ the antighost field and $B^a$ the Nakanishi-Lautrup multiplier. Evidently, the splittings~\noeq{asplit} and \noeq{c.split} induce a corresponding separation of these latter fields into their soft and semi-fast components  according to
\beq
\bc^a = \bcsf^a+\bcso^a;\qquad
\b^a = \bsf^a+\bso^a.
\eeq

Due to the very simple form that the classical equation of motion assumes in the Coulomb gauge [see \1eq{poisson}] the soft gauge fixing function will be chosen to be the Coulomb gauge one
\beq
{\cal F}^a_{\mathrm{soft}}(p)=-i p^i \delta A^a_i(p),
\eeq
while the semi-fast function will be left, at the moment, unspecified.
The ghost-dependent terms in \noeq{sgf} can be computed
by using the BRST transformations in \2eqs{BRST.deltaA}{BRST.semifast}; one obtains in this case
\begin{align}
S_\s{\rm FPG} =&- \int\!\dif^4p\,\t(|p^+|-b\Lambda) \t(\Lambda - |p^+|) \bcsf^a(-p)s{\cal F}^a_{\mathrm{s.fast}}(p)  \nonumber \\
&-\int\!\dif^4p\,
\t(b \Lambda - |p^+|) \bcso^a(-p) p_i^2 \cso^a(p)\nonumber \\
&+\int\!\dif^4p\!\int\!\dif^4q\,  \t(b \Lambda - |p^+|)g f^{abc}  \bcso_a(-p)ip^i \delta A^b_i(q) \cso^c(p-q).
\end{align}

\section{\label{sec:sti}Slavnov-Taylor Identity}

Until now our analysis has been strictly classical.
In order to carry out the quantization of the theory, one needs
a procedure to promote to the quantum level the nonlinear symmetry generated by the BRST operator $s$. The most efficient way to accomplish this  is through the introduction of certain external sources $\Phi^*$ (one for each field $\Phi$ transforming non-linearly under the symmetry being considered)  which describe the renormalization of the composite operators that are bound to appear.  These sources, called {\it antifields}~\cite{Gomis:1994he}, have  opposite statistics with respect to the corresponding field $\Phi$, ghost charge $\mathrm{gh}(\Phi^*)=-1-\mathrm{gh}(\Phi)$, and,  choosing the (mass) dimension of the Faddeev-Popov ghost fields to be zero, dimension\footnote{These conventions ensure that the action has both ghost number as well as canonical dimension zero.} $\mathrm{dim}(\Phi^*)=4-\mathrm{dim}(\Phi)$.

Antifields are then coupled to the composite operators generated by the BRST variation of the corresponding field through the term 
\beq
S_\s{\mathrm{AF}}=\int\!\dif^4x\sum\Phi^*(x)s\, \Phi(x).
\eeq
Then the invariance of the corresponding (background gauge-fixed) tree-level action $\Gamma^{(0)}$ under the BRST symmetry is encoded in the following ST identity
\beq
{\cal S}\hspace{0.025cm}\Gamma^{(0)}\equiv\int\!\dif^4z\left[
\Omega^a_\mu(z)\frac{\delta\Gamma^{(0)}}{\deltam\widehat{A}^a_\mu(z)}+\sum\frac{\delta\Gamma^{(0)}}{\delta\Phi^*(z)}\frac{\delta\Gamma^{(0)}}{\delta\Phi(z)}\right]=0,
\label{st.tree}
\eeq
where the sum goes over all the fields of the model.
If the theory is anomaly free\footnote{Notice that 
the theory in the presence of the background is non-anomalous if and only if it is anomaly-free at zero background. This follows since, as already said, 
$\widehat A_\mu$ and $\Omega_\mu$ form a BRST doublet~\cite{Barnich:2000zw,Quadri:2002nh} and hence they do not alter the cohomology of the gauge theory~\cite{Barnich:2000zw}.}, the same identity holds for the quantum vertex functional $\Gamma$ (that is, for the generator of the one-particle irreducible amplitudes):
\beq
{\cal S}\hspace{0.025cm}\Gamma=0;\qquad\Gamma=\Gamma[\Phi;\widehat A;\Phi^*;\Omega].
\eeq

In the CGC framework the procedure explained above is complicated by the fact that its tree level action
\beq
\Gamma^{(0)}=S_\s{\mathrm{CGC}}+S_\s{\mathrm{GF}}+S_\s{\mathrm{FPG}}+S_\s{\mathrm{AF}},
\label{cgc-tl}
\eeq
involves fields with support in different momentum regions, for which the corresponding antifields have to be defined. Specifically, the splittings of \2eqs{asplit}{c.split} requires the introduction of the semi-fast antifields $a^*$,  $c^*$ and the soft ones $\deltam A^*$, $\delta c^*$.
Thus, BRST invariance of the tree-level CGC action~\noeq{cgc-tl} is expressed through the ST identity
\begin{align}
{\cal S}\hspace{0.025cm}\Gamma^{(0)}\equiv\int\!\dif^4z&\left[\Omega^a_\mu(z)\frac{\delta\Gamma^{(0)}}{\deltam\widehat{A}^a_\mu(z)}+\frac{\delta\Gamma^{(0)}}{\delta{a}^{*a}_\mu(z)}\frac{\delta\Gamma^{(0)}}{\delta{a}^a_\mu(z)}+\frac{\delta\Gamma^{(0)}}{\delta(\deltam{A}^{*a}_\mu(z))}\frac{\delta\Gamma^{(0)}}{\delta(\deltam{A}^a_\mu(z))}\right.\nonumber \\
&+\frac{\delta\Gamma^{(0)}}{\delta(\cso^{*a}(z))}\frac{\delta\Gamma^{(0)}}{\delta(\cso^a(z))}+\frac{\delta\Gamma^{(0)}}{\delta{\csf}^{*a}(z)}\frac{\delta\Gamma^{(0)}}{\delta{\csf}^a(z)}
+\bsf^a(z)\frac{\delta\Gamma^{(0)}}{\delta\bcsf^a(z)}+\bso^a(z)\frac{\delta\Gamma^{(0)}}{\delta(\bcso^a(z))}
\nonumber \\
&\left.+\frac{\delta\Gamma^{(0)}}{\delta(\deltam{\chi}^{*a}(z))}\frac{\delta\Gamma^{(0)}}{\delta(\deltam{\chi}^a(z))}\right]=0,
\label{ST}
\end{align}
which generalizes to the vertex functional $\Gamma$
\begin{align}
{\cal S}\hspace{0.025cm}\Gamma&=0;&\Gamma=\Gamma[a,\deltam A,\csf, \cso,\bcsf,\bcso,\bsf,\bso,\deltam\chi;\widehat A, \widehat\chi;a^*,\deltam A^*,\csf^*,\cso^* \deltam\chi^*;\Omega].
\end{align}
Notice that we have not introduced an antifield for $\bcsf$ and $\bcso$, since they transform linearly under the BRST operator\footnote{The situation is similar to the one discussed in the case of the pair $(\widehat{A},\Omega)$: $(\bcsf,b)$ and $(\bcso, \delta b)$ form BRST doublets, and no antifield is necessary.}.

As has been previously explained, in the CGC framework, one is interested in the correlators of the quantum fluctuations $\deltam\chi$ of the color charge density, once the semi-fast modes $a$ have been integrated out.
Such correlators are therefore one-particle reducible w.r.t. all the semi-fast modes ($a$, $\csf$, $\bcsf$, $\bsf$), and are generated by a new effective action $\widetilde \Gamma[\deltam A, \deltam c, \deltam \bar c, \deltam b,\deltam\chi;\widehat A, \widehat\chi;\deltam A^*,\deltam c^*, \deltam\chi^*;\Omega]$ satisfying an ST identity which differs from~\1eq{ST}.

The effective action $\widetilde{\Gamma}$ can be formally obtained by starting 
from the connected generating functional ${\cal W}$, 
which is the Legendre transform of $\Gamma$ w.r.t. the different fields of the theory; one has
\begin{align}
{\cal W} &= \Gamma + \int\! \dif^4 x \, \sum J_\Phi \Phi;&
J_\Phi &= - (-1)^{\epsilon(\Phi)} \frac{\delta \Gamma}{\delta \Phi}; &
\Phi& = \frac{\delta {\cal W}}{\delta J_\Phi};  &
\frac{\delta {\cal W}}{\delta\Phi^*} &= \frac{\delta \Gamma}{\delta\Phi^*},
\end{align}
 where $J_\Phi$ denotes the source of the quantum field $\Phi$, while $\epsilon(\Phi)$ represents the statistics of the field $\Phi$ ($1$ for anticommuting variables, $0$ for commuting ones).
Notice that $\Phi$ runs on all the quantum fields, including both the soft and the semi-fast modes.

Then, in terms of the connected generating functional ${\cal W}$, the ST identity~\noeq{ST} reads
\begin{align}
\int\!\dif^4z\,\Omega^a_\mu(z)\frac{\delta {\cal W}}{\delta\widehat{A}^a_\mu(z)}&=\int\!\dif^4z\,\frac{\delta {\cal W}}{\delta(\delta{A}^{*a}_\mu(z))}J_{\deltam A^\mu_a}(z)+\int\!\dif^{4}z\,\frac{\delta{\cal W}}{\delta{a}^{*a}_\mu(z)}J_{a^\mu_a}(z)
\nonumber \\
& 
-\int\!\dif^{4}z\,\frac{\delta{\cal W}}{\delta(\cso^{*a}(z))}J_{\cso^a}(z)
-\int\!\dif^{4}z\,\frac{\delta{\cal W}}{\delta \csf^{*a}(z)}J_{\csf^a}(z)&\nonumber\\
&
-\int\!\dif^{4}z\,\frac{\delta{\cal W}}{\delta J_{\bso^a}(z)}J_{\bcso^a}(z)
-\int\!\dif^{4}z\,\frac{\delta{\cal W}}{\delta J_{\bsf^a}(z)}J_{\bcsf^a}(z)
\nonumber \\
&+\int\!\dif^{4}z\,\frac{\delta{\cal W}}{\delta(\deltam{\chi}^{*a}(z))}J_{\deltam\chi^a}(z).
\label{W-STI}
\end{align}
Next, we define $\widetilde{\Gamma}$ by setting to zero each source associated to the fields we want to integrate out; 
this amounts to imposing their equation of motion and, diagrammatically,
to consider amplitudes that are one-particle reducible (1-PR) w.r.t. such fields. We then obtain
\begin{align}
\widetilde\Gamma
=&\left.{\cal W}\right\vert_{{J_a}=J_\csf=J_{\bcsf}=J_{\bsf}=0}+\int\!\dif^4z\,J_{\deltam A^\mu_a}(z)\deltam A^\mu_a(z)+\int\!\dif^4z\,J_{\deltam \chi^a}(z)\deltam \chi^a(z)
\nonumber \\
& + \int\!\dif^4z\,J_{\cso^a}(z)\cso^a(z)
  + \int\!\dif^4z\,J_{\bcso^a}(z)\bcso^a(z)
  + \int\!\dif^4z\,J_{\bso^a}(z)\bso^a(z).
\label{Legendre}
\end{align}
Notice that as we are not taking the Legendre transform w.r.t. $a$, $\csf$, $\bcsf$ and $\bsf$, 
$\widetilde\Gamma$ contains one-particle reducible diagrams with respect to these fields.

Finally, by setting $J_{a}=J_{\csf}=J_{\bcsf}=J_{\bsf}= 0$ in~\1eq{W-STI} and performing afterwards the Legendre transform~\noeq{Legendre}, one finds the modified ST identity
\begin{align}
{\cal S}\hspace{0.025cm}\widetilde \Gamma\equiv\int\!\dif^4z&\left[\Omega^a_\mu(z)\frac{\delta\widetilde \Gamma}{\deltam\widehat{A}^a_\mu(z)}+\frac{\delta\widetilde \Gamma}{\delta(\deltam{A}^{*a}_\mu(z))}\frac{\delta\widetilde \Gamma}{\delta(\deltam{A}^a_\mu(z))}
+\frac{\delta\widetilde \Gamma}{\delta(\cso^{*a}(z))}\frac{\delta\widetilde \Gamma}{\delta(\cso^a(z))}+\bso^a(z)\frac{\delta\widetilde \Gamma}{\delta(\bcso^a(z))}\right.\nonumber \\
&\left.+\frac{\delta\widetilde \Gamma}{\delta(\deltam{\chi}^{*a}(z))}\frac{\delta\widetilde \Gamma}{\delta(\deltam{\chi}^a(z))}\right]=0.
\label{ST-tildeGamma}
\end{align}
Let us emphasize once again that even though this functional equation has the same form as 
the original ST identity in~\1eq{st.tree},
amplitudes in $\widetilde \Gamma$ are not 1-PI w.r.t. the semi-fast modes.

Eq.~(\ref{ST-tildeGamma}) has a rich physical meaning and expresses in compact form
two important properties of the quantized theory: (i) as a consequence of the
BRST symmetry associated with the SU($N$) gauge invariance, at $\Omega_\mu =0$
one obtains the ST identity for the theory of the soft modes in the presence
of a fast background; (ii) by taking
a derivative w.r.t. $\Omega_\mu$ and then setting $\Omega_\mu = 0$, one gets
the quantum-deformed equation of motion for the background fast field $\widehat A_\mu$, which
can be solved by a specific canonical transformation derived in~\cite{Binosi:2012st,Binosi:2012pd}, 
allowing to reconstruct the full background dependence in the quantum theory.
We will analyze each of these properties in more detail in the following two subsections.

\subsection{Gauge invariance}\label{sec:gaugeinv}

By taking a derivative w.r.t. $\cso$ of~\1eq{ST-tildeGamma} and
then setting $\cso$ and $\Omega$, as well as $\bso$, to zero, one finds
\beq
 \int \dif^4z \left[ 
\frac{\delta^2 \widetilde \Gamma}{\delta (\cso_b(x)) 
\delta(\deltam{A}^{*a}_\mu(z))} \frac{\delta\widetilde \Gamma}{\delta(\deltam{A}^a_\mu(z))}
+
\frac{\delta^2 \widetilde \Gamma}{\delta (\cso_b(x))
\delta(\deltam{\chi}^{*a}_\mu(z))} \frac{\delta\widetilde \Gamma}{\delta(\deltam{\chi}^a_\mu(z))} \right]=0.
\label{ward.id}
\eeq
We now notice that
 $\deltam{A}^*$ is coupled to the BRST variation $s\, \delta A$
of~\1eq{BRST.deltaA}, while 
$\deltam{\chi}^*$ couples to the BRST variation $s\, \delta \chi$ 
of~\1eq{sondchi}; thus, neither of them is coupled (at $\Omega = 0$) to $a$ and $c$. 
Since
in $\widetilde \Gamma$ one does not integrate over
$\delta A$ and $\cso$, the Green's functions involving one
soft antifield remain classical,
namely
\begin{align}
& \frac{\delta^2 \widetilde \Gamma}{\delta (\cso_b(x)) 
\delta(\deltam{A}^{*a}_\mu(z))} = \frac{\delta^2 \widetilde \Gamma^{(0)}}{\delta (\cso_b(x)) 
\delta(\deltam{A}^{*a}_\mu(z))} =  \delta^{ab} \partial_\mu \delta^{(4)}(x-z)
+ g f^{acb} \delta A^c_\mu \delta^{(4)}(x-z), \nonumber \\
& \left . \frac{\delta^2 \widetilde \Gamma}{\delta (\cso_b(x))
\delta(\deltam{\chi}^{*a}(z))} \right |_{\Omega = 0} =
 \left . \frac{\delta^2 \widetilde \Gamma^{(0)}}{\delta (\cso_b(x))
\delta(\deltam{\chi}^{*a}(z))} \right |_{\Omega = 0} =
g f^{acb} \left[\wh\chi^c(x) + \delta \chi^c(x)\right] \delta^{(4)}(x-z).
\end{align}
Hence \1eq{ward.id} amounts to the statement
of gauge invariance of the effective action $\widetilde \Gamma$.
It is very important to notice that this result holds
{\em irrespectively of the gauge choice for the semi-fast modes} ${\cal F}_{\mathrm{s.fast}}$: that is, gauge invariance follows as a consequence of the ST identity
(\ref{ST-tildeGamma}) and of the 
decomposition between semi-fast and soft modes, once the appropriate BRST symmetry,
induced by this splitting, is taken into account.

\subsection{\label{sec:qcor}Background (quantum-corrected) equations of motion}

The identity~(\ref{ST-tildeGamma})
can be further simplified if we restrict our attention to the dependence on the background field for amplitudes
involving only external $\delta A$ and/or $\delta \chi$ legs. This is clearly the case we want to focus on, as  the $\deltam\chi$ correlators generated by $\widetilde\Gamma$ are those that will be eventually identified with the momenta of the updated weight function $W_{b\Lambda}[\chi]$, {\it i.e.}, the original $W_\Lambda[\chi]$ after a one step quantum evolution. Setting $J_{\delta c^a} =
J_{\delta \bar c^a} = 0$ in~\1eq{W-STI}, and taking again the
Legendre transform one gets in this case the {\it reduced} identity
\beq
{\cal S}\hspace{0.025cm}\widetilde\Gamma\equiv\int\!\dif^4z\left[\Omega^a_\mu(z)\frac{\delta\widetilde\Gamma}{\deltam\widehat{A}^a_\mu(z)}+\frac{\delta\widetilde\Gamma}{\delta(\deltam{A}^{*a}_\mu(z))}\frac{\delta\widetilde\Gamma}{\delta(\deltam{A}^a_\mu(z))}+\frac{\delta\widetilde\Gamma}{\delta(\deltam{\chi}^{*a}(z))}\frac{\delta\widetilde\Gamma}{\delta(\deltam{\chi}^a(z))}
\right]=0.
\label{ST-final}
\eeq

Taking a functional differentiation of~\1eq{ST-final} w.r.t. the source~$\Omega$, and setting $\Omega=0$ afterwards, yields 
\beq
\frac{\delta\widetilde\Gamma}{\deltam\widehat{A}^a_\mu(x)}=-\int\!\dif^4z\left[\frac{\delta\widetilde\Gamma}{\delta\Omega^{a}_\mu(x)\delta(\deltam{A}^{*b}_\nu(z))}\frac{\delta\widetilde\Gamma}{\delta(\deltam{A}^a_\mu(z))}
+\frac{\delta\widetilde\Gamma}{\delta\Omega^{a}_\mu(x)\delta(\deltam{\chi}^{*b}(z))}\frac{\delta\widetilde\Gamma}{\delta(\deltam{\chi}^b(z))}
\right].
\label{EoM-fundam}
\eeq
This is a fundamental equation for the EFT at hand, as it encodes how quantum fluctuations will modify the classical equation of motion~\noeq{maxrho}. Such knowledge is evidently required in order to be able to reconstruct the gluon fields generated by the new color charge density. 

Indeed, the first term in~\1eq{EoM-fundam} controls the (gauge-dependent) deformation of the classical background-quantum splitting~\noeq{asplit} induced by quantum corrections~\cite{Binosi:2011ar,Binosi:2012pd,Binosi:2012st}; 
the second term fixes instead the functional dependence of the background $\wh{A}$ on the color charge density $\deltam\chi$, once quantum corrections are taken into account. 
This result is completely general, for it does not rely on the specific
form of the action chosen, the gauge-fixing adopted for the
semi-fast modes, or even the order of approximation used
while carrying out the one-step quantum evolution. 

\section{\label{sec:def} Gauge invariant analysis of the deformation functions}

\begin{figure}   
\includegraphics[scale=0.7]{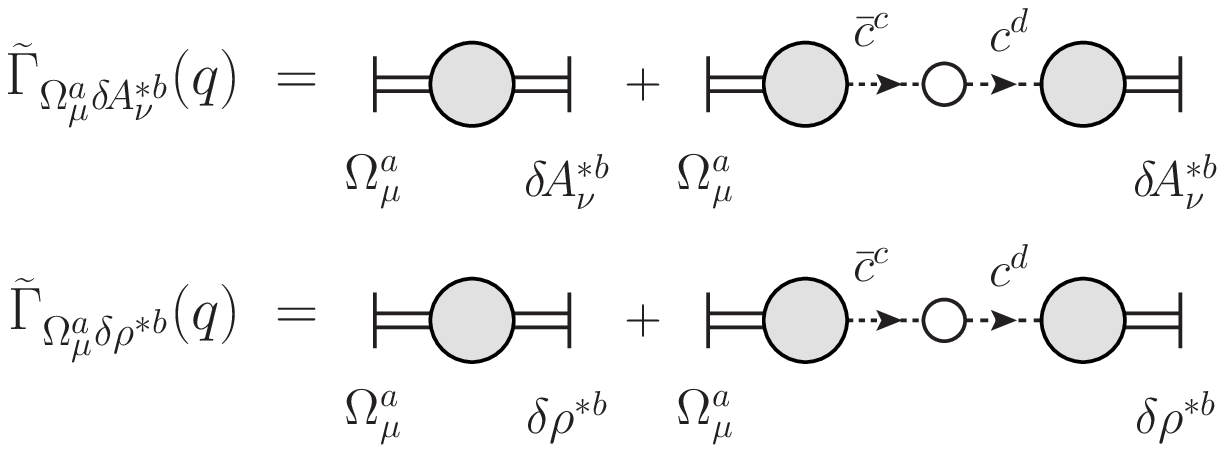} 
\caption{\label{G_Omega_deltaAstar}The deformation functions~$\widetilde{\Gamma}_{\Omega\,\deltam A^*}(q)$ and~$\widetilde{\Gamma}_{\Omega\,\deltam \rho^*}(q)$. As these functions are 1-PR w.r.t the semi-fast fields, on top of the 1-PI diagram one finds a 1-PR contributions proportional to the ghost propagator.}   
\end{figure}

Consider now the deformation functions 
$\widetilde\Gamma_{\Omega\, \deltam A^*}$ and $\widetilde\Gamma_{\Omega\, \deltam \chi^*}$ explicitly appearing in~\1eq{EoM-fundam},
defined by 
\beq
\widetilde\Gamma_{\Omega^a_\mu(x)\, \deltam A^{b*}_\nu(y)} \equiv
\frac{\delta^2 \widetilde\Gamma}{\delta \Omega^a_\mu(x) \delta  (\deltam A^{b*}_\nu(y))} \, , \qquad
\widetilde\Gamma_{\Omega^a_\mu(x)\, \deltam \chi^{b*}(y)} \equiv
\frac{\delta^2 \widetilde\Gamma}{\delta \Omega^a_\mu(x) \delta  \chi^{b*}(y)} \, , 
\eeq
and let us analyze their behavior.

To begin with observe that, as shown in~\fig{G_Omega_deltaAstar},  due to the quantum numbers of the source $\Omega$ and the antifields $\deltam A^*$ and $\deltam \chi^*$  there are only two contributions: the 1-PI term and a 1-PR graph in which the semi-fast ghosts are exchanged. In addition, while the couplings of the source $\Omega$ depend on the BRST variation of the (semi-fast) gauge fixing $s{\cal F}_{\mathrm{s.fast}}$, the antifields $\deltam A^*$ and $\deltam \chi^*$ couple only to the BRST variations of the corresponding fields, which, in turn, do not depend on the choice of  ${\cal F}_{\mathrm{s.fast}}$, being fixed by symmetry requirements only.

This being said, let us start with $\widetilde\Gamma_{\Omega\, \deltam A^*}$. As shown in~\1eq{BRST.deltaA} the antifield $\deltam A^*$ couples at most to a soft ghost and a soft gluon. As $\widetilde\Gamma$ is constructed by integrating out the semi-fast modes
(so that there cannot be internal soft lines in the $\widetilde \Gamma$-amplitudes),
 then both the 1-PI as well as the 1-PR terms are zero. Thus one is led to the result
\beq
\widetilde{\Gamma}_{\Omega^\mu_a\deltam A^{*\nu}_b}(x,y)=0,
\label{r1}
\eeq
and, since at no point we have assumed the background field to be zero, taking any number $n$ of functional derivatives w.r.t. the background field yields 
\beq
\widetilde{\Gamma}_{\Omega^\mu_a\deltam A^{*\nu}_b\widehat{A}^{\rho_1}_{c_1}\cdots \widehat{A}^{\rho_n}_{c_n}}(x,y,z_1,\dots,z_n)=0.
\label{r2}
\eeq

The analysis of the deformation function $\widetilde\Gamma_{\Omega\, \deltam \chi^*}$ is similar. Again one observes that the couplings of the antifield $\deltam\chi^*$ are dictated by the BRST variation~\noeq{sondchi} alone; that shows in turn that there are no couplings to any semi-fast mode. Thus again the 1-PR term is zero while the 1-PI diagram is confined at tree-level, as, contrary to the previous case,~\1eq{sonwhchi} generates a tree-level coupling $\Omega\,\deltam\chi^*$. Thus one finds 
\begin{align}
&\widetilde{\Gamma}_{\Omega^\mu_a \delta\chi^*_b}(x,y)  = \Gamma^{(0)}_{\Omega^\mu_a \delta \chi^*_b}(x,y) = - \delta^{\mu+}\bm \nabla^2_{\bx} \delta^{(4)}(x-y),\nonumber \\
&\widetilde{\Gamma}_{\Omega^a_\mu\,\delta\chi^{*b}\widehat{A}^{\rho_1}_{c_1}\cdots \widehat{A}^{\rho_n}_{c_n}}(x,y,z_1,\dots,z_n)=0,
\label{r3}
\end{align}
with the second relation obtained from the first one by taking $n$ functional differentiations w.r.t. the background field.
Notice that the time dependence, induced by the Wilson line
involving $\delta A^-$, has been reabsorbed into the definition
of the $\chi$ charge in Eq.~(\ref{chi-charge}),
thus leading to  Eqs.~(\ref{r3}).

Once again, the results established in~\3eqs{r1}{r2}{r3} are independent from the form of the semi-fast gauge-fixing function ${\cal F}_{\mathrm{s.fast}}$, as for deriving them we never had to resort to any special property of the $\Omega$ source couplings. Thus the vanishing of the deformation function $\widetilde\Gamma_{\Omega\, \deltam A^*}$ and the fact that $\widetilde\Gamma_{\Omega\, \deltam \chi^*}$ is confined at tree-level represent gauge invariant statements. As a result, one has that, independently of the semi-fast gauge fixing for the $a$ modes,~\1eq{EoM-fundam} further simplifies to 
\beq
\frac{\deltam\widetilde\Gamma}{\delta\widehat{A}^a_\mu(x)}=-\int\!\dif^4z\,\frac{\delta\widetilde\Gamma^{(0)}}{\delta\Omega^{a}_\mu(x)\delta(\deltam{\chi}^{*b}(z))}\frac{\delta\widetilde\Gamma}{\delta(\deltam{\chi}^b(z))}.
\label{EoM-fundam-GI}
\eeq
This latter equation represents the full equation of motion for the background field $\widehat{A}$ when the semi-fast quantum fluctuations are integrated out; we once again stress that the deformation function $\Gamma_{\Omega\,\deltam\chi^*}$ remains completely classical and background independent.

\subsection{Physical consequences}
 
The vanishing of the $\widetilde\Gamma_{\Omega\, \deltam A^*}$   deformation function (together with all its background field derivatives) implies that the classical background-quantum splitting in \1eq{asplit} is not deformed after the one step quantum evolution. 
Therefore the identification of $\delta A$ with the
soft mode and of $\wh{A}$ with the fast 
component of the gluon field is not spoiled by the quantum evolution.

This also clarifies an important conceptual point in the consistency of the separation of gluon modes carried out in the CGC framework. Indeed, the expansion of the path-integral over the semi-fast modes [see Eq.(\ref{appseffdef})] is not performed
around a stationary point of the action, as $\wh{A} + \delta A$ 
is not a solution of the classical Yang-Mills equation of
motion~\noeq{maxrho}.
Therefore in general one expects a shift, induced by quantum corrections, of the classical field configuration. 
However, such a shift would be proportional to $\widetilde\Gamma_{\Omega\, \deltam A^*}$~\cite{Binosi:2011ar,Binosi:2012pd,Binosi:2012st} and therefore is absent in the CGC effective field theory.

Finally, \1eq{EoM-fundam-GI} yields a relation between the correlators of the quantum fluctuations $\deltam\chi$ of the color charge density and the background field $\wh{A}$ once the semi-fast modes $a$ have been integrated out.
Consider for example the case of the two-point background sector.
By taking first a derivative w.r.t. $\wh{A}$ and then w.r.t. $\deltam\chi$~\1eq{EoM-fundam-GI} gives
\begin{align}
\widetilde \Gamma_{\widehat{A}_a^\mu \widehat{A}_b^\nu}(x_1,x_2)&=-
\int\!\dif^4z\,\widetilde \Gamma^{(0)}_{\Omega_a^\mu \, \deltam \chi^*_c}(x_1,z)\widetilde \Gamma_{\wh{A}_b^\nu \deltam \chi_d}(x_2,z),
\nonumber \\
\widetilde \Gamma_{\widehat{A}_a^\mu \deltam\chi_b}(x_1,x_2)&=-
\int\!\dif^4z\,\widetilde \Gamma^{(0)}_{\Omega_a^\mu \, \delta \chi^*_c}(x_1,z)\widetilde\Gamma_{\deltam\chi_b  \deltam \chi_d}(x_2,z),
\end{align} 
where all the 2-point functions are to be evaluated at non-zero background. Substituting the second equation into the first one gives the final relation
\beq
\widetilde \Gamma_{\widehat{A}_a^\mu \widehat{A}_b^\nu}(x_1,x_2)=
\int\!\dif^4y\!\int\!\dif^4z\,\widetilde \Gamma^{(0)}_{\Omega_a^\mu  \deltam \chi^*_c}(x_1,y)\widetilde \Gamma^{(0)}_{\Omega_b^\nu  \deltam \chi^*_d}(x_2,z)\widetilde \Gamma_{\delta \chi^c  \deltam \chi^d}(y,z).
\label{2point}
\eeq
By using~\1eq{r3} one therefore obtains the final relation (recall that $\alpha\equiv A^+$)
\beq
\widetilde \Gamma_{\alpha^a \alpha^b}(x_1,x_2)=
\int\!\dif^4y\!\int\!\dif^4z\,\bm{\nabla}_{{\bx}_1}^2\delta(x_1-y)\bm{\nabla}_{{\bx_2}}^2\delta(x_2-y)\widetilde \Gamma_{\delta \chi^a \, \delta \chi^b}(y,z).
\label{2ptalpha}
\eeq

It is instructive to study~\1eq{2ptalpha} at tree-level and compare it with the relations between the correlators fixing the initial conditions for the CGC evolution. By projecting~\1eq{2ptalpha} at order zero in the loop expansion one finds
\beq
\widetilde \Gamma^{(0)}_{\alpha^a \alpha^b}(x_1,x_2)=
\int\!\dif^4y\!\int\!\dif^4z\,\bm{\nabla}_{{\bx}_1}^2\delta(x_1-y)\bm{\nabla}_{{\bx_2}}^2\delta(x_2-y)\widetilde \Gamma^{(0)}_{\delta \chi^a \, \delta \chi^b}(y,z).
\label{2ptalpha_tl}
\eeq
We take now, as a concrete example, a weight function corresponding to the simple Gaussian of the MV model~\cite{Iancu:2002xk} for a large nucleus with atomic number $A \gg 1$:
\beq
W_\s{A}[\chi] = {\cal N} \exp \left\{ -\frac{1}{2} \int\dif^3x \, \frac{\chi^a(\vec{x}) \chi^a(\vec{x})}{\lambda_\s{A}(x^-)} \right\}.
\eeq
In such case, one has (see Eq.~(2.57) of~\cite{Iancu:2002xk})
\begin{align}
\langle \alpha^a(\vec{x}) \alpha^b(\vec{y}) \rangle_\s{A}^{-1}  & =
\delta^{ab} \delta(x^- - y^-) \gamma^{-1}_\s{A}(x^-,x_\perp - y_\perp);&
\gamma_\s{A}(x^-,k_\perp) &= \frac{1}{k^4_\perp} \lambda_\s{A}(x^-).
\label{2pt.corr.mv}
\end{align}

Hence, one can identify the inverse of the propagator
$\langle \alpha^a \alpha^b \rangle_\s{A}^{-1}$ with the 2-point function $\widetilde \G^{(0)}_{\alpha^a \alpha^b}$ and similarly, upon splitting the field $\chi=\widehat\chi+\delta\chi$ as custom in the background field method, $\lambda_\s{A}^{-1}(x^-)$ with $\widetilde \G^{(0)}_{\delta \chi \delta \chi}$.
Since the splitting of the color sources is trivially given by (\ref{chisplit}),
we arrive at the natural identification
\beq
W_\s{A}[\chi] = {\cal N} \exp \left\{-\widetilde \G[\chi] \right\} 
\eeq
where it is understood that the r.h.s. is evaluated at zero
gauge fields (we only look at the correlators of the color sources).

Let us now project~\1eq{2ptalpha} at the one loop level ({\it i.e.}, at the level of approximation of the CGC,  that is $\alpha_s\ln1/b$). In this case one obtains the relation between the correlators of the background fields and those of the color charges $\deltam \chi$ that play a crucial role in establishing the JIMWLK equation.

In order to make contact with the notation of~\cite{Hatta:2005rn}, one has to carry out the following identifications, corresponding to the obvious extension of the classical analysis described above:
\begin{align}
\label{identifications}
\widetilde \rho&\leftrightarrow\widehat\chi;&
\delta\widetilde\rho&\leftrightarrow\deltam\chi;&
\langle{\mathrm T}\left[\delta\alpha\, \delta\alpha\right]\rangle^{-1}&\leftrightarrow\widetilde \Gamma^{(1)}_{\alpha \alpha};&
\langle{\mathrm T}\left[\delta\widetilde\rho\, \delta\widetilde\rho\right]\rangle^{-1}&\leftrightarrow\widetilde \Gamma^{(1)}_{\deltam\chi \deltam\chi}.
\end{align}
Then, the inversion of~\1eq{2ptalpha} yields
\beq
\langle{\mathrm T}[\delta\alpha^a(x_1)\, \delta\alpha^b(x_2)]\rangle=
\int\!\dif^4y\!\int\!\dif^4z\,\bm \Delta(x_1-y)\bm \Delta(x_2-y)\langle{\mathrm T}[\delta\widetilde\rho^a(y)\, \delta\widetilde\rho^b(z)]\rangle,
\eeq
which should be compared with Eq. (4.30) of~\cite{Hatta:2005rn}.

\section{\label{sec:evol}Evolution equations}

Eq.~(\ref{2ptalpha}) shows that, to all orders in the loop expansion,
the relation between the background field and the
color charge density remains classical. This in turn implies
that the non-trivial effects of the one-step quantum evolution
can be described by studying the evolution of the 
correlators of the color sources.

Within the framework we have introduced so far, one can obtain the  JIMWLK equation through the (rather natural) requirement that the correlators of the induced source $\delta\widetilde{\rho}$ appearing on the lhs of the first of Eqs.~(4.27) of~\cite{Hatta:2005rn} coincide with the correlators of the quantum fluctuations $\delta\chi$ in our quantum EFT.
Setting $\tau=\alpha_s\ln1/b$ one has (we will suppress for the moment all color and Lorentz indices)
\bea
\frac{\partial}{\partial\tau}\langle T\left[\delta\chi(x_1)\cdots\delta\chi(x_n)\right]\rangle&=&\frac{\partial}{\partial\tau}\int\!{\cal D}[\delta\chi]\,\delta\chi(x_1)\cdots\delta\chi(x_n)\exp\left[\tau\Delta S_\s{\mathrm{eff}}+\cdots\right]\nonumber\\
&=&\int\!{\cal D}[\delta\chi]\,\delta\chi(x_1)\cdots\delta\chi(x_n)\Delta S_\s{\mathrm{eff}}\exp\left[\tau\Delta S_{\mathrm{eff}}+\cdots\right],
\label{2star}
\eea
where the dots indicate $\tau$-independent terms, and one has used the fact that the $\tau$ dependence at leading logarithmic order is a factor in front of $\Delta S_{\mathrm{eff}}$. 
The latter is certainly true when calculating the effective action with the undressed gluon propagator; 
moreover, from the analysis given in Appendix~\ref{app:jimwlk} we see that this will be true even when using the background dressed propagator.

Consider now the expansion of the effective action at $\delta A^-=0$; one has \beq
\Delta S_\s{\mathrm{eff}} = \sum_{m=0}^\infty\frac1{m!}\int_y\underbrace{\frac{\delta^m\Delta S_{\mathrm{eff}}}{\delta(\delta A^-(y_1))\cdots\delta(\delta A^-(y_m))}}_{\Gamma_m(y_1,\dots,y_m)}
\delta A^-(y_1) \cdots \delta A^-(y_m),
\label{star}
\eeq
where we have defined $\int_y=\int\!\dif^4y_1\cdots\int\!\dif^4y_m$;
notice, in addition, that the coefficient functions $\Gamma_m(y_1,\dots,y_m)$ are background-dependent.

At the relevant order in the eikonal approximation 
$\delta \chi$ is coupled to $\delta A^-$ through a bilinear vertex 
in $S_\s{W}$ in Eq.~(\ref{sw2}). Therefore, at this order of approximation, in the CGC effective field theory
(where no quantum integration is carried out over $\delta A^-$),
each external $\delta A^-$ leg can be converted
into a $\delta \chi$ leg. Physically this means that
 $\delta A^-$ plays the role of the source of $\delta \chi$.
By taking this fact into account 
the expansion in the r.h.s. of Eq.~(\ref{star}) 
yields
\bea
\frac{\partial}{\partial\tau}\langle T\left[\delta\chi(x_1)\cdots\delta\chi(x_n)\right]\rangle&=&\sum_{m=0}^\infty\frac1{m!}\int_y\!\Gamma_m(y_1,\dots,y_m)\int\!{\cal D}[\delta\chi]\,\delta\chi(x_1)\cdots\delta\chi(x_n)\delta\chi(y_1)\cdots\delta\chi(y_m)
\nonumber \\
&\times& \exp\left[\tau\Delta S_\s{\mathrm{eff}}+\cdots\right].
\eea

Let us next introduce the notation
\beq
{\cal G}^{n;m}(x_1,\dots,x_n;y_1,\dots,y_m)=\langle T\delta\chi(x_1)\cdots\delta\chi(x_n)\delta\chi(y_1)\cdots\delta\chi(y_m)\rangle,
\eeq
so that the previous equation reads
\beq
\frac{\partial}{\partial\tau}{\cal G}^{n;0}(x_1,\dots,x_n)=\sum_{m=0}^\infty\frac1{m!}\int_y\Gamma_m(y_1,\dots,y_m){\cal G}^{n;m}(x_1,\dots,x_n;y_1,\dots,y_m).
\label{3star}
\eeq
Now, recall that the field $\delta\chi$ describes the fluctuations of the charge density induced by quantum corrections; therefore the corresponding correlators will be associated to the momenta of the classical probability distribution (which incorporates semi-fast quantum corrections). Thus one has the identification
\beq
{\cal G}^{n;m}(x_1,\dots,x_n;y_1,\dots,y_m)\equiv\frac{\delta^{n+m}W}{\delta(\delta\chi)(x_1)\cdots\delta(\delta\chi)(x_n)\delta(\delta\chi)(y_1)\cdots\delta(\delta\chi)(y_m)},
\eeq
so that~\1eq{3star} will read
\begin{align}
\frac{\partial}{\partial\tau}\frac{\delta^{n}W}{\delta(\delta\chi)(x_1)\cdots\delta(\delta\chi)(x_n)}&=\sum_{m=0}^\infty\frac1{m!}\!\int_y\Gamma_m(y_1,\dots,y_m)\times\nonumber \\
&\times\frac{\delta^{n+m}W}{\delta(\delta\chi)(x_1)\cdots\delta(\delta\chi)(x_n)\delta(\delta\chi)(y_1)\cdots\delta(\delta\chi)(y_m)},
\end{align}
which for $n=0$ gives the following evolution equation
\beq
\frac{\partial}{\partial\tau}{W}=\sum_{m=0}^\infty\frac1{m!}\int_y\!\Gamma_m(y_1,\dots,y_m)\frac{\delta^{m}{W}}{\delta(\delta\chi)(y_1)\cdots\delta(\delta\chi)(y_m)}.
\label{eveq}
\eeq

The derivation of this equation equation relies only on the assumption that the $\tau$ dependence factors out; the form of  $\Delta S_{\mathrm{eff}}$ is not needed. Therefore~\1eq{eveq} is valid for both  the JIMWLK and its BFKL limit, the difference between the two cases being given by the form that the correlation functions $\Gamma_m(y_1,\dots,y_m)$ assume. 

By comparing Eq.~(\ref{eveq}) with Eq.~(\ref{star}) one gets back the well-known result that the effective Hamiltonian can be obtained by replacing $\delta A^-(x)$ in the effective action with the differential operator $\frac{\delta}{\delta (\delta\chi)(x)}$.  However, a remark is in order here.
In the eikonal approximation the couplings between $\delta A^-$ and 
both $\widehat \chi$ and $\delta \chi$ are the same, since they are obtained
from $\int\!\dif^4x \, \delta A^-(x) \chi(x)$ after the splitting
$\chi = \widehat \chi + \delta \chi$. Therefore one can safely
replace everywhere in Eq.~(\ref{eveq}) $\delta \chi$ with $\widehat \chi$ and get
the customary form of the Hamiltonian evolution for $W$.
This is in agreement with the previously stated prescription that
the Green's functions
of the classical $\widehat \chi$ source, generated by $W$,
coincide with the correlators of the quantum fluctuations $\delta \chi$
in the effective field theory of the CGC.
Notice however that in general one cannot dispose of $\delta \chi$ all together (and work only with $\widehat \chi$), since $\delta \chi$ is required in order
to formulate the reduced ST identity (\ref{ST-final}), which holds to all
orders in the perturbative expansion.

What we have achieved here is separating the derivation of the general form of the evolution equation (which is dictated by the symmetries of the theory alone) from the calculation of $\Delta S_{\mathrm{eff}}$, which  is rather a result of the approximations one would like to introduce due to the particular regime he is interested in. 
The detailed evaluation of $\Delta S_{\rm eff}$ for the JIMWLK equation is given in Appendix~\ref{app:jimwlk}.

\section{Conclusions}\label{sec:concl}

In this paper we have clarified the role played by the fundamental BRST symmetry of the QCD action in constraining the form of the high-energy evolution equation the theory can give rise to. In particular, we have concentrated on the EFT of the CGC, and achieved a complete separation between the general features of evolution equations, that only depend on the symmetry content of the theory, from the specific aspects related, {\it e.g.}, to the choice of the gauge for the semi-fast modes or the particular approximation used in the computation of the EFT (one-loop) action.

The crucial enabling step has been the identification of the correct BRST symmetry of the CGC theory holding after the gluon field has been separated into its fast, semi-fast and soft components. Once this has been done, the corresponding ST identity encoding at the quantum level the (classical) BRST invariance of the action can be written down. 

A plethora of results then naturally follows. To begin with, the gauge invariance of the EFT (after the one-step quantum evolution) is a direct consequence of the mere existence of this identity. As a second result, one is able to prove that the classical Yang-Mills equations of motion are not deformed by the quantum corrections induced when integrating out the semi-fast field $a$. This implies that the classical description of the CGC at the new scale $b\Lambda$ in terms of a modified weight function $W_{b\Lambda}[\rho]$, with the same equations of motion holding at the scale $\Lambda$, is indeed consistent. This is a crucial ingredient in the derivation of the evolution equations, and it is 
remarkable that it can be derived on the basis of symmetry arguments
only. 
Finally, one can prove that the exact form of the evolution equation is determined by the BRST symmetry alone; the approximations made in the calculation of the effective action account instead for which of the various evolution equations known in the literature ({\it e.g.}, the JIMWLK or its BFKL limit) one is using. It should be noticed that in deriving all the aforementioned results, at no point we have fixed the gauge for the semi-fast modes $a$, which, as a matter of fact, has been left unspecified\footnote{It is however convenient in the CGC theory to fix the background Coulomb gauge for the fast background $\widehat{A}$, due to the particularly easy form that the backgound equation of motion assumes in this case.}.

Besides being a framework in which the theory of QCD high energy evolution equations possibly admits its most rigourous formulation, the methods introduced here might help in going beyond the approximations usually employed. In particular, a more general evolution equation than the one presented in~\noeq{eveq} can be derived by using algebraic techniques~\cite{Piguet:1995er}, thus dropping the assumption of linearity in $\tau$. This very interesting research direction clearly deserves further investigation.   

\acknowledgments
We would like to thank Edmond Iancu for useful and stimulating discussions. 

\appendix
\section{\label{app:jimwlk}Effective action for the JIMWLK equation}

In this Appendix we give a ``sketch'' of the derivation of the JIMWLK Hamiltonian. The total field $A_{\mu}$ in the CGC is split according to \eqn{asplit} and the goal is to construct an effective action $S_{\rm eff}$ quadratic in $\delta A^-$ and to all orders in $\widehat{A}_{\mu}$, by integrating the semi-fast modes $a_{\mu}$ having longitudinal momenta $k^+$ such that 
$b\Lambda \ll |k^+| \ll \Lambda$. Therefore we define
 \beq\label{appseffdef}
 \exp(\rmi S_{\rm eff}) =
 \int_{b \Lambda}^{\Lambda}
 \mcal{D}a \exp \left( \rmi S \right).
 \eeq
$S$ is the sum of the CGC action $S_{\rm CGC}$, 
given by \eqn{action}, and the gauge-fixing and 
Faddeev-Popov part $S_{\rm GF+ FPG}$ in Eq.(\ref{sgf}).

$S_{\rm CGC}$ generates the classical field equations in the presence of a static source $\rho(\vec{x})$ when $\delta A^{\mu}=0$, as in~\eqn{maxrho}. The background field $\widehat{A}^{\mu}$ in \eqn{asplit} can be determined by the solution to the Poisson equation as we explain now. In the light-cone gauge only the transverse components are non-zero, i.e.~$\widehat{A}_{\mu} = \delta_{\mu i} \widehat{A}_i$, and therefore we have
 \beq
 \label{appwharho}
 \mcal{D}_{\nu} F^{\nu +} =
 -\mcal{D}_{i}\, \del^{+} \widehat{A}^i = \rho(\vec{x}).
 \eeq
Here we have made use of the fact that $F^{-+} = - \del^{+} \delta A^{-} \simeq 0$, which comes about because 
$\delta A^{-}$ contains modes with very small longitudinal momenta and thus its variation with $x^{-}$ is very slow. The solution to the classical equation \eqref{appwharho} is
 \beq\label{appwhafu}
 \widehat{A}^i = \frac{\rmi}{g}\,
 U(\vec{x})\, \del^{i} U^{\dag}(\vec{x}),
 \eeq
with the Wilson line $U^{\dag}$ given by
 \beq\label{appwilson}
 U^{\dag}(\vec{x}) =
 {\rm P} \exp \bigg[
 \rmi\, g \int_{-\infty}^{x^-}
 \dif z^- \alpha^a(z^-,\bm{x})\, T^a
 \bigg].
 \eeq
In the above we have defined $\alpha \equiv \widehat{A}^+$ which is the only non-vanishing component of the background gauge field in the Coulomb gauge and satisfies
 \beq\label{apppoisson}
 \nabla^2_{\bm{x}} \alpha(\vec{x}) = -\chi(\vec{x}),
 \eeq

with $\chi(\vec{x})$ the corresponding (static) source in the Coulomb gauge. 
Notice that in the JIMWLK approximation we set the 
$W$ rotation equal to 1,  because the rest of the calculation gives already the dominant $(\delta A^-)^2$ dependence.

The choice of the Coulomb gauge for the soft modes and of the
light-cone gauge for the semi-fast ones uniquely fixes $S_\s{\rm GF+ FPG}$.
Since in the light-cone gauge the ghosts decouple, they can be neglected
while performing the one-step quantum evolution. Moreover, in order
to derive the effective action required to obtain the JIMWLK Hamiltonian, the
Green's functions of the Nakanishi-Lautrup field $\bso$ and
of the soft ghosts and antighosts $\cso, \bcso$ are not needed
and thus one can simply take $S=S_\s{\rm CGC}$.

Now we expand the action $S$ around
$A^{\mu}_0 \equiv \delta^{\mu i} \widehat{A}^{i} + \delta^{\mu-} \delta A^{-}$ and, in view of the Gaussian integration to follow, to second order in the semi-fast modes $a^{\mu}$, that is
  \beq\label{appsexpansion}
  S = S_0 + \frac{\delta S}{\delta A^{i}}\, a^{i} +
  \frac{1}{2}\,a^{\mu}\, G_{\mu\nu}^{-1}\, a^{\nu}.
  \eeq
It should be noticed that the expansion is not around the solution
of the Yang-Mills equation of motion $\widehat{A}$, rather around
$A_0$. I.e. one is not expanding around a stationary point  of
the action and thus, in general, one would expect that quantum corrections
will deform the classical background solution.
However in the CGC this does not occur, as a consequence of the stability
of the background equation of motion (\ref{EoM-fundam-GI}).

The propagator $G_{\mu\nu}$ is in the presence of only the background field $\widehat{A}^i$, since the field $\delta A^{-}$ can be set equal to zero to the order of accuracy. The coefficient of the linear term may be written as
 \beq\label{appdsda}
 \frac{\delta S}{\delta A^{i}}  =
 \mcal{D}_{\nu}F^{\nu i} =
 \mcal{D}_{j }F^{ji} + \mcal{D}^{+} F^{-i} 
 + \mcal{D}^{-} F^{+i}
 \simeq 2 \mcal{D}^+ F^{-i},
 \eeq
where we have used the fact that $F^{ij}$ is a pure gauge, i.e.~$F^{ij}=0$, and also the approximate equality 
$\mcal{D}^{-} F^{+ i} \simeq  \mcal{D}^+ F^{-i}$. The latter is due to the fact that $\del^+ \delta A^{-} \simeq 0$ as justified earlier. Now it becomes straightforward to perform the integration over the semi-fast modes $a^{\mu}$ to obtain the change of the effective action. It is given by a four-dimensional double integral, more precisely
 \beq
 \label{appseff1}
 \Delta S_{\rm eff} =
  -\,\frac{1}{2} 
 \int \dif^4 x\, \dif^4 y\,
 [2 \mcal{D}^+ F^{-i}(x)] \,
 G^{i j}(x,y)\, 
 [2 \mcal{D}^+ F^{-i}(y)].
 \eeq
So far we have been working in the light-cone gauge,
where ghost modes decouple and therefore can be neglected
when integrating out the soft modes.
In explicit computations this is a major simplification, that
favours the light-cone gauge  with respect to 
other gauge choices for the $a_\mu$ fields, in which ghosts
do not decouple any more.

Since \eqn{appseff1} has a gauge invariant form, it will be more convenient to calculate it in the Coulomb gauge. Even though the propagator is an important element of the calculation, since it also contains the logarithmic enhancement, here we shall just give its final form. It reads \cite{Hatta:2005rn}
 \beq
 \label{appprop}
 G^{i j}(x,y) = -
 \frac{\rmi\, \Delta \tau}{4\pi}\,
 \delta^{ i j}\,
 \delta^{(2)}(\bm{x} - \bm{y})
 \big[\Theta(x^- - y^-) 
 U^{\dag}_{x^ - y^-}(\bm{x})
 + \Theta(y^- - x^-) 
 U_{y^-x^-}(\bm{x}) \big],
 \eeq
with $\Delta \tau = \ln(1/b)$ representing as usual the differential enhancement in the longitudinal phase space and where the Wilson line $U^{\dag}$ is given by a similar expression to the one in \eqref{appwilson}, but with the lower limit replaced by $y^-$. Using the fact that the propagator satisfies $\mcal{D}^{+}(x) G^{ij}(x,y)=0$,
it is not hard to show that the integrand in
\eqref{appseff1} is a total derivative with respect to both
$x^-$ and $y^-$ and thus the integration is determined by the surface terms. Furthermore, the propagator is
independent of the light-cone time and we can immediately integrate over $x^+$ and $y^+$. Using $F^{-i} = -\del^{i} \delta A^{-}$ and defining for notational simplicity $A^{-}(\vec{x}) = \int \dif x^+ \delta A^{-}(x)$ we arrive at
 \begin{align}
 \hspace*{-0.2cm}
 \label{appseff2}
 \frac{\Delta S_{\rm eff}}{\Delta \tau} =
 \frac{\rmi}{2\pi}
 \int &\dif^2 \bm{x}\,
 \big[
 \del^{i} A^{-}(\infty,\bx)\,
 \del^{i} A^{-}(\infty,\bx) +
 \del^{i} A^{-}(-\infty,\bx)\,
 \del^{i} A^{-}(-\infty,\bx)
 \nn
 &- \del^{i} A^{-}(\infty,\bx)\,
 U^{\dag}(\bm{x})\,
 \del^{i} A^-(-\infty,\bx)
 - \del^{i} A^{-}(-\infty,\bx)
 U(\bx)\,
 \del^{i} A^{-}(\infty,\bx)
\big],
 \end{align}
where now the Wilson lines are as in \eqn{appwilson}), but with the integration extending over the whole longitudinal axis. The analysis in Sect.~\ref{sec:evol} suggests that the evolution Hamiltonian can be obtained via the replacement 
$A^{-}(\vec{x}) \to -\rmi \delta/\delta \chi(\vec{x})$. Then, by making use of the Poisson equation \eqref{apppoisson} we can express $\delta/\delta \chi(\vec{x})$ in terms of $\delta/\delta \alpha(\vec{x})$. Since the functional
derivatives will act at the end-points, $+\infty$ or 
$-\infty$, of the Wilson lines we find that
 \beq\label{appdda}
 \frac{\delta}{\delta \alpha^a(-\infty,\bx)}
 =\frac{\delta}{\delta \alpha^b(\infty,\bx)}\,
 \big[U^{\dag}(\bm{x})\big]^{ba} =
 \big[U(\bm{x})\big]^{ab}
 \frac{\delta}{\delta \alpha^b(\infty,\bx)},
 \eeq
and thus we can express the functional derivatives at $x^- = -\infty$ in terms of those at $x^- = \infty$. This brings us to the ``standard'' form of the JIMWLK Hamiltonian
    \beq\label{apphk}
    H = \frac{1}{(2\pi)^3}
    \int
    \dif^2\bm{u}\,
    \dif^2\bm{v}\,
    \dif^2\bm{z}\,
    K_{\bm{u}\bm{v}\bm{z}}\,
    \frac{\delta}{\delta \alpha^a_{\bm{u}}(\infty)}
    \big[ 1
    +U_{\bm{u}}^{\dag} U_{\bm{v}}
    -U_{\bm{u}}^{\dag} U_{\bm{z}}
    -U_{\bm{z}}^{\dag} U_{\bm{v}}
    \big]^{ab}
    \frac{\delta}{\delta \alpha^b_{\bm{v}}(\infty)},
    \eeq
where we have denoted the dependence on transverse coordinates with an index and with the kernel $K_{\bm{u}\bm{v}\bm{z}}$ given by
  \beq\label{appk}
  K_{\bm{u}\bm{v}\bm{z}}=
  \frac{(\bm{u}-\bm{z})}{(\bm{u}-\bm{z})^2}
  \cdot
  \frac{(\bm{v}-\bm{z})}{(\bm{v}- \bm{z})^2}.
  \eeq


\end{document}